\documentclass[aps,pre,twocolumn,showpacs,superscriptaddress,longbibliography]{revtex4-1}
\usepackage{graphicx}
\usepackage{dcolumn}
\usepackage{bm}
\usepackage{amssymb}
 \usepackage{xcolor}
 
\begin{document}
\title{Kinetic and hydrodynamic regimes in multi-particle-collision 
dynamics \\ of a one-dimensional fluid with thermal walls
}
\author{Stefano Lepri}
\affiliation{Consiglio Nazionale delle Ricerche, Istituto dei Sistemi Complessi via Madonna del piano 10, I-50019 Sesto Fiorentino, Italy}
\affiliation{Istituto Nazionale di Fisica Nucleare, Sezione di Firenze, via G. Sansone 1, I-50019 Sesto Fiorentino, Italy} 
\author{Guido Ciraolo}
\affiliation{CEA, IRFM, F-13108 Saint-Paul-lez-Durance, France}
\author{Pierfrancesco Di Cintio}
\affiliation{Dipartimento di Fisica e Astronomia and CSDC, Universit\'a di Firenze, 
via G. Sansone 1, I-50019 Sesto Fiorentino, Italy}
\affiliation{Istituto Nazionale di Fisica Nucleare, Sezione di Firenze, via G. Sansone 1, I-50019 Sesto Fiorentino, Italy}
\author{Jamie Gunn}
\affiliation{CEA, IRFM, F-13108 Saint-Paul-lez-Durance, France}
\author{Roberto Livi}
\affiliation{Dipartimento di Fisica e Astronomia and CSDC, Universit\'a di Firenze, 
via G. Sansone 1, I-50019 Sesto Fiorentino, Italy}
\affiliation{Istituto Nazionale di Fisica Nucleare, Sezione di Firenze, via G. Sansone 1, I-50019 Sesto Fiorentino, Italy}
\affiliation{Consiglio Nazionale delle Ricerche, Istituto dei Sistemi Complessi via Madonna del piano 10, I-50019 Sesto Fiorentino, Italy}
\date{\today}
\begin{abstract}
We study the non-equilibrium steady-states of a 
one-dimensional ($1D1V$) fluid in a finite space region of length $L$.  
Particles interact among themselves by multi-particle collisions and are in 
contact with two thermal-wall heat reservoirs, located at the boundaries of the region. 
 After an initial ballistic regime, we find a crossover from a normal (kinetic) transport regime to
an anomalous (hydrodynamic) one, above a characteristic 
size $L_*$. We argue that $L_*$ is proportional to the 
cube of the collision time among particles.
Motivated by the physics of emissive divertors in fusion
plasma, we consider  the more general case of thermal walls
injecting particles with given average (non-thermal) velocity. 
For fast and 
relatively cold particles, short systems fail to establish
local equilibrium and display non-Maxwellian distributions
of velocities. 
\end{abstract}
\maketitle
\section{Introduction}

Statistical systems driven away from equilibrium by external agents,
like concentration and/or thermal gradients, forced flows etc. 
are ubiquitous in nature and in technological applications. 
Especially in far-from-equilibrium regimes 
simulation of model systems is a basic tool to gain insight
on the fundamental properties of the system under study. 
Molecular dynamics is the most natural approach, but methods based 
on effective stochastic processes may be a computationally convenient
alternative.
In particular, the Multi-Particle Collision (MPC) dynamics can be 
considered as a mesoscale simulation
method where particles undergo stochastic collisions.
The implementation was originally proposed by Malevanets and
Kapral \cite{Malevanets1999, kapral08} and consists of two distinct 
stages: a free streaming and a collision one. 
Collisions occur at fixed discrete time intervals, 
and space is discretized into cells that 
define the collision range. The MPC dynamics is a useful 
tool to investigate concrete systems like polymers in solution, 
colloidal fluids etc. but also to 
address fundamental problems in statistical physics and
in particular the effect of external sources.

In the present work, we investigate the transport property of 
a one-dimensional gas driven off-equilibrium by different 
reservoirs, modeled as thermal walls. This general setup has been considered
very often in the literature\citep{LLP03,DHARREV} with several 
possible applications \citep{Lepri2016}.
For the discussion that follows, it is worth recalling that
the prediction of Nonlinear Fluctuating Hydrodynamics \cite{VanBeijeren2012,Spohn2014} for 1D systems, constrained by three global conservation laws, is that transport is generically anomalous and 
belongs to the Kardar-Parisi-Zhang (KPZ) dynamical universality
class. This is confirmed by 
other approaches including hydroynamics \cite{Narayan02,Delfini06}, 
kinetic theory \cite{Pereverzev2003,Nickel07,Lukkarinen2008}  and
many numerical simulations \cite{Lepri03,Wang2011,Liu2014}.
However, for finite systems some deviations are found and it is 
thus of interest to clarify which are the different transport 
regimes (see e.g. Refs.\cite{Benenti2020,Lepri2020} for 
an up-to-date account). In this framework, simple kinetic
model are an unvaluable testbed.

Besides the general theoretical motivations, our interest in also in 
application of the technique in plasma physics 
\cite{DiCintio2015,DiCintio2017}. 
We wish to investigate the MPC method  as a tool to study
nonequilibrium properties of fusion plasma which are
subject to large temperature gradients.
This is a relevant issue in the modeling of plasma transport in the direction parallel
to the magnetic field in the edge of magnetic fusion devices where
hot plasma regions are connected to a wall
component. In this case a significant temperature gradient 
along the field line sets in between the hot region, that acts as a
heat source, and the colder plasma region at the wall, that acts as
a sink. Deviations from the classical Fourier law have been observed
(see for example \cite{Stangeby00, Krasheninnikov20} and references therein) together with
transitions from strongly collisional to collisionless regimes, exhibiting different heat conduction features. 
We want to point out that accounting for these kinetic effects by a fluid model of
heat transport is of primary importance for implementing realistic
and efficient hydrodynamic simulations. In this perspective, what contained in this work 
can be considered as a worthwhile evolution along the line of 
simplified 1D fluid  models, already adopted for studying thermal transport in plasmas \cite{Bond1981,Bufferand2010,Bufferand2013}. 
 For the sake of brevity, we adopt for such 
model the label $1D1V$ used sometimes in the literature, meaning that we consider particles 
constrained on the line with one velocity component along it.

Besides this, in the edge plasma of magnetic fusion devices several phenomena, like for example neutral beam injection are related with the injection of hot and/or cold plasma particles. We refer for example to hot sources related to external heating of the plasma but also to particular regimes with emissive divertor targets on the wall due to the high temperatures associated with steady state loads, as expected for example on next step large fusion devices \cite{kocan2015impact,gunn2017surface}. 
Here, we will thus attempt to model this situation, by generalizing 
the usual thermal wall method to include the possibility of 
injecting particles with an average non-thermal velocity. To 
our knowledge this case has not been treated in the statistical
physics literature and it has thus an interest by itself. 

The outline is as follows.
In Section \ref{sec:model} we recall the definition of the fluid dynamics and the 
thermal walls. Section \ref{sec:kin} discusses the crossover from the so 
called kinetic (diffusive) regime to the anomalous, hydrodynamic one.
The effect of general thermal walls is illustrated in Sec. \ref{sec:general}
with reference to the specific case of shifted Maxwell-Boltzmann distributions.

\section{1D1V MPC with thermal walls}
\label{sec:model}

For a detailed description of the MPC simulation scheme we address the reader to references \cite{DiCintio2015,DiCintio2017}. 
Here, we just summarize the basic ingredients. We consider an ensemble of 
$N$ point particles with equal masses $m$ 
located in a finite space region, $[0,L]$, partitioned into $N_c$ 
(fixed) cells of size $a$, $L=N_ca$. The density of particles, $d=N/L$, is 
fixed, while all physical quantities, without prejudice of generality, are set to dimensionless  units, namely $a=1$, $m=1$ and (for the isolated system)
$E_c=1$, the latter denoting the total kinetic energy of the fluid of particles.

The MPC collision is performed at regular times steps, separated by a constant time interval of 
duration $\delta t$:
the velocity of the $j$-th particle in the $i$-th cell is changed according to the update rule $v_{j,{\rm old}} \to v_{j,{\rm new}}=a_iw_j+b_i$,
where $w_j$ is randomly sampled by a thermal distribution at the cell temperature $T_i$, while $a_i$ and $b_i$ are cell-dependent
parameters, determined by the condition of total momentum and total energy conservation in the cell \cite{DiCintio2015}. 
After the collision all particles freely propagate, i.e.
\begin{equation}
x_j(t+\delta t) = x_j(t) + v_j(t) \, \delta t \, .
\label{free}
\end{equation}
and they may move to  a new cell.

The non-equilibrium state is imposed by two thermal 
walls \cite{Banavar1998} acting at the edges of the space region:
any particle crossing  the boundaries of
the interval $[0,L]$  is re-injected in the opposite direction
with a random  velocity extracted form the Maxwell-Boltzmann distributions 
\begin{equation}
p_0(v)=  \frac{v}{T_0} e^{-v^2/2T_0} ;\quad 
p_L(v)=- \frac{v}{T_L} e^{- v^2/2T_L} \, .
\end{equation}
Re-injected particles propagate with their new velocities for 
the remaining time up to the next collision. 
Since arbitrarily large values of $v$  are admitted, a particle may
reach the opposite boundary during this time, although such a process, in practice,  becomes extremely unlikely, i.e. negligible,  for 
sufficiently large values of $L$, considering that in the adopted dimensionless units  the typical value
of $v$ is ${\mathcal O}(1)$.

The heat fluxes at the two boundaries, $Q_0$ and $Q_L$, respectively,
are computed as the average  kinetic
energy exchanged by particles interacting with the walls \cite{LLP03,DHARREV}.
The relaxation of the fluid to a steady state,
i.e.  $Q_0\approx -Q_L=Q$, is monitored  
by controlling the convergence of the running-time  averages of the fluxes.
The thermal conductivity is defined as  $K = Q/(\Delta T/L)$,
where $\Delta T=T_0-T_L$. 
Typically, in order to speed up the relaxation to a steady state, the initial 
velocities of the particles in cell $i$ are randomly extracted from a  Maxwell-Boltzmann distribution
at temperature $T_i =  T_L+(T_0-T_L)i/L$, i.e. according to the linear temperature profile compatible with
the Fourier's law.

The relevant kinetic
parameters of the MPC protocol are 
the particle mean free path $\ell$, which,  in an uniform system, 
is proportional
to the thermal velocity of the fluid $v_T = \sqrt{T}$, in formulae
$
\ell \sim v_T \delta t  \,,
$
and the thermal conductivity, that is expected to be proportional
to the typical kinetic energy of the fluid, in formulae $K = Cv_T\ell \sim v_T^2 \delta t$, $C$
being the specific heat at constant volume.

We conclude this section by making the reader aware of the main advantage of MPC simulations, with
respect to those typically performed in nonlinear lattice models. The possibility of observing crossover between
different transport regimes in the latter class of models depends on the fine tuning of various simulation parameters,
e.g. energy density, integration time step, coupling with the thermal reservoirs, and all that are strongly model-dependent,
because the crossover effects are ruled by the lifetime of typical nonlinear excitations, influencing the stationary transport process.
As we show in what follows, in MPC simulations one can control different regimes essentially by a single natural parameter, the collision
time $\delta t$, which is straightforwardly related to the mean-free path of the fluid particles, weak and strong collisional
regimes corresponding to large and small values of $\delta t$, respectively.

\section{From kinetic to hydrodynamic regimes}
\label{sec:kin}

The kind of stationary heat transport phenomenon one is dealing with is characterized by the scaling of $Q$  with 
the system size $L$. Diffusive transport, i.e. Fourier's law, yields  $Q \propto L^{-1}$. For
the $1D1V$ MPC,  based on general theoretical arguments \cite{VanBeijeren2012,Spohn2014}
we expect that transport would be anomalous and belonging to the KPZ universality 
class, as indicated by equilibrium correlation
functions \citep{DiCintio2015}. 
In the nonequilibrium setup used in the present work this would imply that
\begin{equation}
Q \propto L^{-2/3} \qquad L\to \infty \, .
\label{Q23}
\end{equation}
However, this asymptotic anomalous regime may be eventually attained going through a crossover from a standard kinetic,
i.e. diffusive, regime \cite{Zhao2018,Miron2019,Lepri2020}. This scenario can be rationalized
by assuming that the flux is made of two contributions
\begin{equation}
Q=Q_N+Q_A \, ,
\label{qaqn}
\end{equation}
where  $Q_N$ is the "normal" part of the flux
that can be written as  
\begin{equation}
Q_N = \frac{q}{r+L/\ell}
\label{qn}
\end{equation}
(with $q$ and $r$ being suitable parameters) while $Q_A$ is the anomalous part that, 
for large enough values of $L$, should scale as
\begin{equation} 
 Q_A\sim \left(\frac{b}{L}\right)^\frac{2}{3} \, ,
\end{equation}
and $b \sim {\mathcal O}(1)$ is a characteristic length \cite{Lepri2020}.
 Formula (\ref{qn}) is suggested by kinetic theory to account for the crossover from 
initial ballistic regime to a diffusive one, see e.g. Ref. \citep{Aoki01}
and Section 3.4 in Ref. \citep{LLP03}. 
It implies that 
up to system lengths of the order of the mean free path $L\sim \ell$
transport is essentially ballistic, with a flux independent of
$L$ while for $L\gg \ell$, $Q_N \sim \frac{\ell}{L}$.

According to the above formulae, for large enough $L$ one can define a crossover length $L_*$ from the normal to the anomalous regime when $Q_N\approx Q_A$, i.e.
\begin{equation}
L_*\sim \left(\frac{\ell^3}{b^2}\right) \sim v_T^{3}\, \delta t ^{3}.
\end{equation}
 Altogether, upon increasing $L$ at fixed $\ell$, one should 
see a first ballistic regime followed by a kinetic (diffusive) one until 
eventually the asymptotic hydrodynamic regime is attained.
It should be however remarked, that the different regimes
are observable only provided that the relevant length scale are 
widely separated, in such a way that the range of scales 
between the ballistic and anomalous regimes is large enough 
i.e. $\ell \ll L_*$.

\begin{figure}
\includegraphics[width=0.45\textwidth]{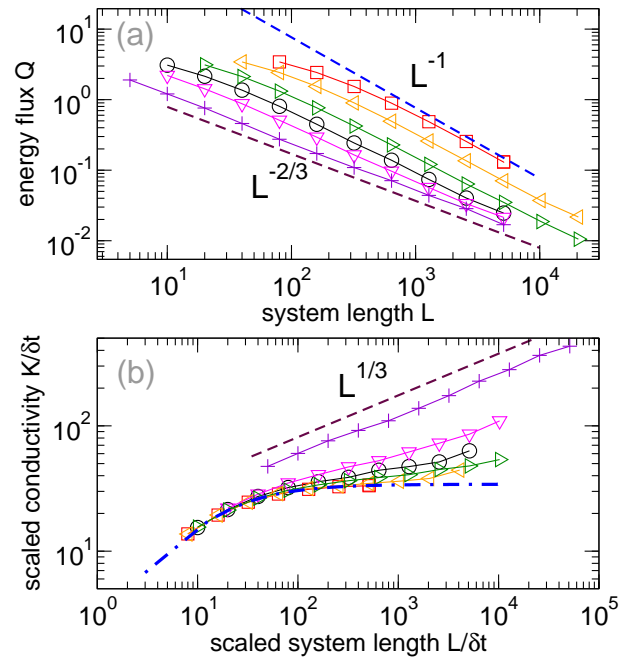}
\caption{Simulations of MPC fluid with density $d=5$, 
$T_0=4$, $T_L=2$. (a) Scaling of the energy fluxes $Q$ as a function of the 
size $L$ for increasing collision times $\delta t=0.1, 0.5, 1.0, 2.0,
5.0$ and $10$ (bottom to top); the upper and the lower dashed lines  correspond
to the scaling of normal and anomalous transport, respectively.
(b) Scaling of the heat conductivity $K=QL/\Delta T$ with the system size $L$;
the dot-dashed blue line is a fit of the data with $\delta t=10$
with the functional form $34.2x/(12.2+x)$, 
 see Eq.(56) in Ref.\cite{LLP03}}.
\label{fig:fscaling}
\end{figure}

The numerical results reported in Fig.\ref{fig:fscaling}(a) provide a clear 
confirmation of this scenario. The energy flux $Q(L)$ displays 
a crossover from diffusive to anomalous scaling upon decreasing
the collision time $\delta t$. 
In Fig.\ref{fig:fscaling}(b)
we also show that, for low collisionality and small $L$, the thermal conductivity $K$, obtained for different values of $\delta t$,
tend to collapse onto  a single curve after rescaling 
$K$ and $L$ by $\delta t$, or, equivalently, by $\ell$ . 
This confirms that in the kinetic regime $\ell$ is the only
relevant scale as expected.

A more refined analysis of the data, supporting {\sl Ansatz}
(\ref{qaqn}), can be obtained by estimating the two contributions
in Eq.(\ref{qaqn}). This is accomplished by fitting rescaled 
conductivity data with the largest $\delta t$ via the functional
form $\frac{A L }{B+L}$ suggested by Eq. (\ref{qn}), with $A,B$ being fitting parameters.

\begin{figure}
\includegraphics[width=0.45\textwidth]{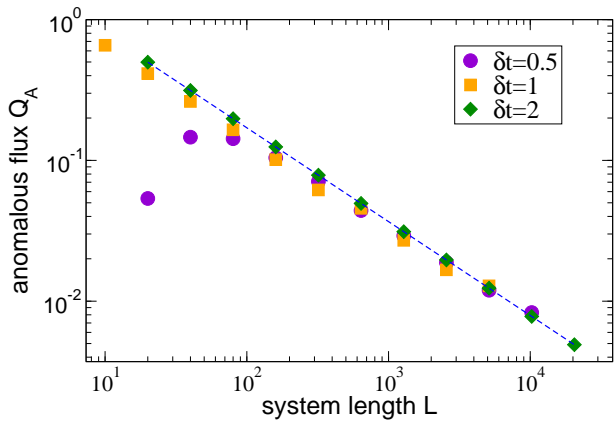}
\caption{Scaling of the anomalous component of the 
energy currents $Q_A$ for $d=5$, 
$T_0=4$, $T_L=2$ and different collision times. The flux $Q_A$
is computed subtracting the normal component $Q_A$ as detailed
in the text. The dashed line is a best-fit of the data
with $\delta t=2$, yielding a decay $Q_A\propto L^{-0.66(5)}$.}
\label{fig:qa}
\end{figure}

As seen in Fig.\ref{fig:fscaling}(b) (dash-dotted line) 
the fitting is very accurate meaning that the anomalous
contribution to the conductivity is negligible 
over the considered length range.
From the fitting function, one can obtain straightforwardly the 
best estimate of $Q_N$ and thus the anomalous part as $Q_A=Q-Q_N$.
As shown in Fig.\ref{fig:qa}, 
a power law fit of $Q_A$ indicates an excellent agreement with the expected 
scaling Eq.(\ref{Q23}) on a remarkably large range of about three decades
in $L$. To our knowledge, this one of the 
most solid numerical confirmations of the predictions of fluctuating hydrodynamics.
We also note, that the characteristic length $b$ is found to be independent 
of the collision parameter.

The difference between the two regimes can be further appreciated also in the 
temperature and density profiles computed as averages
of $v_i^2$ and of the number of particles in each one of the $N_c$ cells.
In order to deal properly with the large $L$ limit, we plot the
temperature and the density profiles, $T(x)$ and $n(x)$, respectively,
as a function of rescaled variable 
$x=i/L$, where $i$ is the cell index. This representation allows one to conclude also that 
local equilibrium is attained, since the product 
$T(x) n(x)$ is $x$-independent, as expected for an ideal
gas (data not shown).

\begin{figure}
\includegraphics[width=0.5\textwidth]{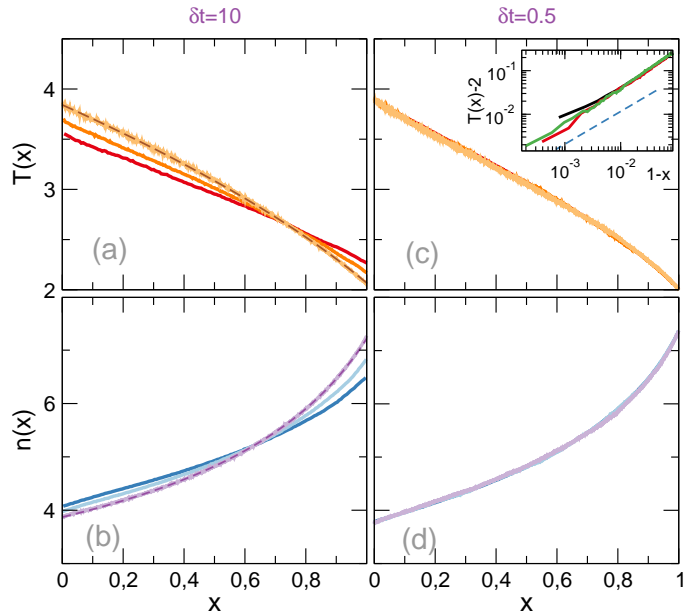}
\caption{Simulations of MPC fluid with density $d=5$, 
$T_0=4$, $T_L=2$. Kinetic regime ($\delta t=10$): temperature (a)  and 
density (b)  profiles for $L=160,320,1280$
(bottom to top);  dashed lines are fits with the function in Eq.(\ref{tk})
yielding $T_0=3.9$, $T_L=2.0$. Hydrodynamic ($\delta t=0.5$) regime:
temperature (c)  and 
density (d)  profiles for $L=640,1280,2560$;
the inset in panel (c) shows the singularity at the right edge of the temperature profile,
the dashed line corresponding to a power-law behavior 
$T(x) \sim (1-x)^{3/4}$ (see
text for details).
}
\label{fig:profi}
\end{figure}

In Fig.\ref{fig:profi}
we report these profiles  for two different values of the collision time
and different sizes $L$. 
For $L<L_*$ the temperature profiles are in good agreement with
the prediction of Fourier's law, 
with thermal conductivity given by the kinetic theory, namely
$K(T)\propto T$. In fact, solving the stationary heat equation 
$(K(T)T')'=0$ on the domain $[0,L]$
with imposed boundary temperature $T(0)=T_0$ and  $T(1)=T_L$
one obtains
\begin{equation}
T(x) = \left[ T_L^{2} + (T_0^{2}-T_L^{2})x\right]^{1/2}.
\label{tk}
\end{equation}
As shown in Fig.\ref{fig:profi}(a) this functional form
accounts for the measured shapes for large enough $L$, apart the
temperature jumps at the edges, that are a typical manifestation of boundary impedance effects.
A similar situation was found also for the HPG model \cite{Dhar2001,Grassberger02}. 
Also, considerations of the same type
apply to the density profiles in Fig.\ref{fig:profi}(b), in view of the relation $T(x) n(x) \approx constant $.

In the hydrodynamic regime, i.e. $L>L_*$, 
the temperature profiles exhibit instead 
the typical signatures of anomalous transport, signaled by singularities at the boundaries
(see  Fig.\ref{fig:profi}(c,d)).
For instance, the inset in Fig.\ref{fig:profi}(c) shows that 
$T(x)\sim(1-x)^\mu$ for $x\to 1$. This property is traced 
back to the fact that steady state profiles should satisfy
a fractional heat equation \citep{Lepri2011b}. The parameter
$\mu$ has been termed meniscus exponent $\mu$, as
it described the characteristic curvature close to the 
edges. In the simulations here  $\mu\approx 3/4$, a value already found 
for other models with reflecting boundary conditions \cite{Lepri2011b,Kundu2019}.

\section{General thermal walls}
\label{sec:general}

Having in mind the possibility of simulating  the injection of hot and/or cold particles into a fusion plasma by  the MPC protocol, in this Section we describe how one can cope with this task by introducing suitable and
more general boundary conditions.
Within the adopted thermal wall scheme this can be achieved by changing the 
distributions of re-injected particles at one boundary of the space region. For instance,
it can be assumed that the distribution $p_0$ on the left boundary is replaced by
a function of the general form \cite{Ianiro1985}
\[
 p_0=v H(v)\quad . 
\]
By imposing the condition 
 $\int_0^\infty vH(v)dv=1$, we ensure that the net flux of particles vanishes  \cite{Ianiro1985}.
This distribution can be characterized by two parameters,
namely the typical average velocity of the injected particles  
$$
v_0 = \int_0^\infty v^2 H(v)dv
$$
and the typical velocity variance
$$
(\Delta v)^2 = \int_0^\infty v^3 H(v)dv - v_0^2 \, ,
$$
that measures the source spread
\footnote{Compare with the case of the a standard Maxwellian bath with $V=0$ 
$H(v) = \beta e^{-\beta v^2/2}$ using the integrals
$v_0 = \int_0^\infty v^2 H(v)dv=\sqrt{\pi/2\beta}$ and 
$\int_0^\infty v^3 H(v)dv = 2/\beta $
one finds $v_T=\sqrt{0.429../\beta}$. Thus 
$v_0$ and $\Delta v$ are both proportional to $\sqrt{T}$
and cannot be changed independently.}.

As a specific example, here we consider a shifted Maxwellian distribution
\begin{equation}
H(v) = H_0(V) e^{- (v-V)^2/2T_0} \, ,
\label{hv}
\end{equation}
parametrized by the velocity $V$, where $H_0$ is a suitable normalization
factor. Shifted Maxwellian sources are implemented in kinetic models to account for drift speed of plasmas \cite{gunn2005kinetic}. Manifestly, both $v_0$ and $\Delta v$ are functions of $V$ and $T_0$.  
In the following simulations, we generate random velocities drawn from (\ref{hv}) 
and use  $V$ and $T_0$ as control parameters, reporting also 
the corresponding values of $v_0$ and $\Delta v$ for comparison
\footnote{In order to generate random deviates with the above distribution
one can take $v=\sqrt{X^2+Y^2}$, where $X,Y$ are iid Gaussian variables with 
nonzero average $A$. In practice, $A$ is fixed to some preassigned
value and the resulting $V,v_0$ are measured by fitting the sampled
distribution Eq. (\ref{hv}).}.

\begin{figure}
\includegraphics[width=0.45\textwidth]{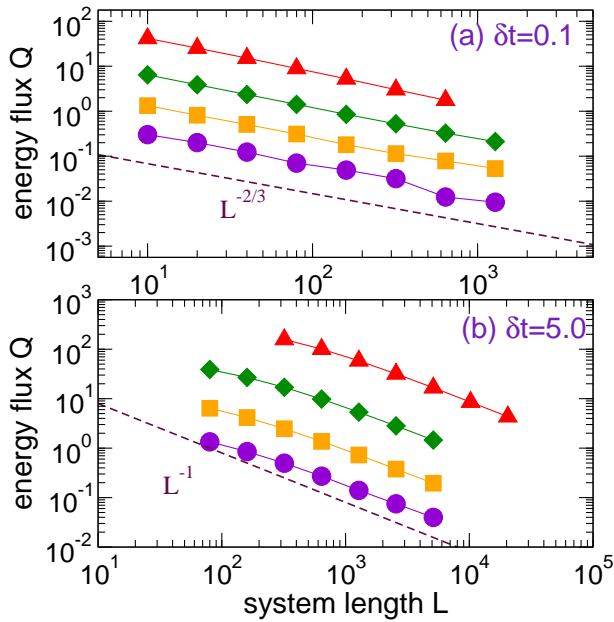}
\caption{Simulations with thermal wall given by Eq.(\ref{hv}) on
the left-hand boundary for different values 
of the parameter $V$, $T_0=T_L=3$ $d=5$;
(a) $\delta t=0.1$,(b) $\delta t=5.$ $V=1,2,4,8$ from bottom
to top. The dashed lines denote the 
asymptotic decay corresponding to normal and anomalous transport. }
\label{fig:fvel}
\end{figure}

In Fig.\ref{fig:fvel} we present the results of  MPC simulations 
for different values of $\delta t$ and $V$. We first of all 
note that, 
even setting equal temperatures at the thermal walls, i.e. $T_0=T_L$, the system is out of equilibrium 
and there is a net flux of energy and particles. For what concerns the scaling with of the flux with the size
$L$, the effect of these thermostats is pretty similar to the 
standard case discussed in the previous Section.
As shown in   Fig.\ref{fig:fvel}a, in the case of 
relatively large collisionality the anomalous scaling, Eq.(\ref{Q23}) is
attained. For weak collisionality, Fig.\ref{fig:fvel}b the 
diffusive scaling is found in the considered range. Although not
observed in the data, we expect the same crossover described 
above beyond the hydrodynamic scale. 
In both cases, as expected, one observes that the magnitude of the 
current increases upon increasing the speed of the injected particles.
The above results are not obvious, and indicate that regardless
of the nature of the baths the overall scenario is mostly dictated
by bulk properties.

\begin{figure*}
\includegraphics[width=0.9\textwidth]{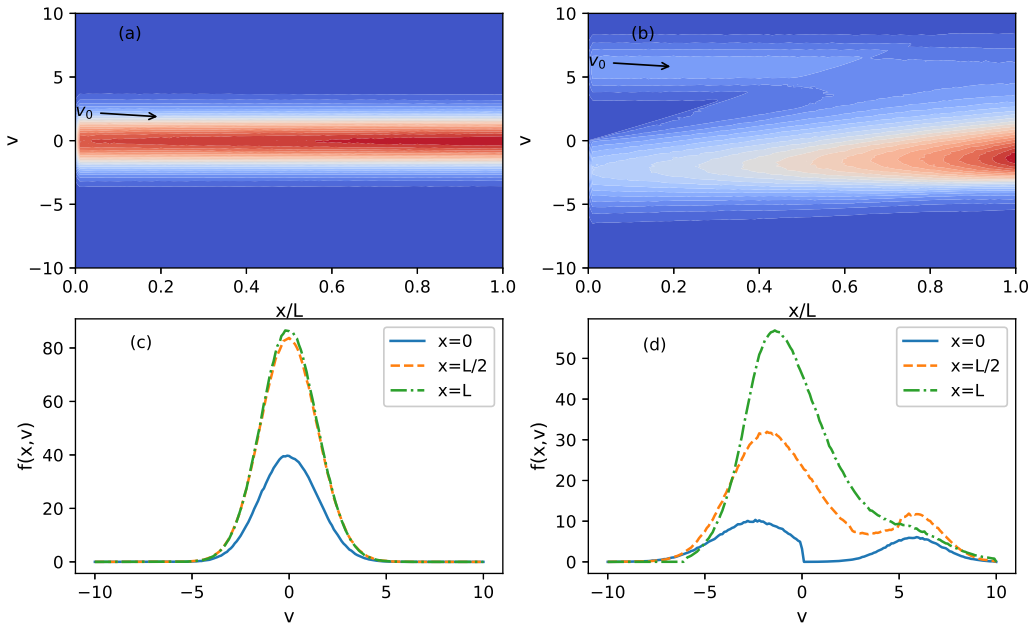}
\caption{Simulations with thermal wall defined by Eq. (\ref{hv}). 
(a,b) Particle distributions $f(x,v)$ in
the steady state   and (c,d) their 
cuts at different positions; the arrows indicate the typical injection 
velocity $v_0$;
$T_0=2$ $T_L=2$, $L=160$, density $d=5$, $\delta t=10$
Left panels (a,c): Injection of slow particles ($V=0.5$), $v_0=1.87$  
$\Delta v=1.0$. Right panels (b,d): Injection of fast and cold particles ($V=4$), 
$v_0=5.81$ $\Delta v=1.24$
Note that even close to the right bath, $x=L$ the 
velocity distribution significantly deviates from a 
Maxwellian.}
\label{fig:fvx}
\end{figure*}

To better appreciate the effect of this kind of thermal wall
we measured the position-dependent steady state distributions 
$f(x,v)$ in phase space for a  moderate collisional parameter 
$\delta t=10$, see Fig.\ref{fig:fvx}. 
We consider two cases: 
\begin{enumerate}
\item the typical entry speed  $v_0$ is of the same order of 
the thermal velocity $v_0 \sim \Delta v = \sqrt{T_0}$ (left
panels in Fig.\ref{fig:fvx}). In this case
the situation is very similar to the standard thermal bath:
temperature and density gradients are formed, while velocity
distributions are Maxwellian even relatively close to the 
left boundary, thus indicating that local equilibrium has been
achieved pretty fast.

\item fast, i.e. "cold", particles are injected, i.e.
the typical entry speed  $v_0$ is larger than $\Delta v$  
(right panels in Fig.\ref{fig:fvx}). In this case one observes that the beam
of particles entering from the left wall propagates, while experiencing
the effect of the interactions 
with the cloud of almost-thermalized particles
only in the center of the space region. 
As a result, the velocity distributions are strongly non-Maxwellian 
and remain double-humped in the first half of the 
space region. 

\end{enumerate}

Adopting a kinetic point of view, we expect that 
the typical size of the region in which the fluid is not 
in local equilibrium would be of the order of mean free path 
$\ell$. For the simulations above, we checked 
that upon increasing $L$ such a region remains indeed 
finite. On the other hand, in anomalously conducting systems lack 
of thermal equilibrium close to the sources is found \cite{Delfini10}
and we cannot exclude that the same occurs here for larger
systems.

\begin{figure*}
\includegraphics[width=0.9\textwidth]{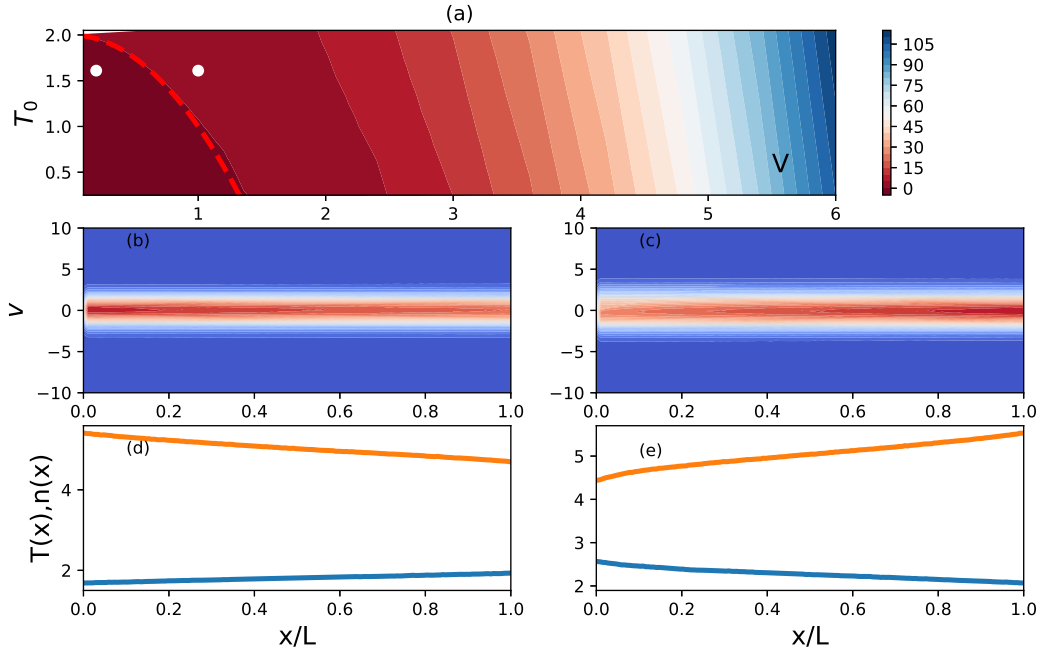}
\caption{Energy current reversal: (a) Contour plot of the flux $Q$ as a function
of the thermal wall parameters $V$ and $T_0$; 
Simulations with other parameters fixed 
$T_L=2$, $L=160$, $d=5$, $\delta t=10$. 
The dashed red line is the curve $T_0=2-V^2$, the energy 
flux is negative on the left side of it (see text for details).
(b,c) the velocity distribution
functions $f(x,v)$ and  (d,e) temperature and density profiles
(lower and upper curves respectively).
Panels (b,d) and (c,e) are for the parameter values
corresponding to $T_0=1.6$ and $V=0.2,1$ respectively
(see the two white dots in the upper panel)
showing change in the sign of the gradients.
}
\label{fig:inv}
\end{figure*}

To conclude this Section, we briefly discuss the dependence of 
fluxes and profiles on the thermal wall parameters. 
In the top panel of Fig. \ref{fig:inv} we plot the 
contour levels of the energy flux $Q$ as a function 
of $V$ and $T_0$ keeping the other parameters fixed.
As remarked above, $Q$ increases with $V$ as intuitively
expected. However there is an interesting feature occurring
for small enough $T_0$ and $V$. For $T_0<T_L$ and $V$ small,
injected particles are still relatively slow 
and thermalise quickly so energy mostly flows from right to 
left, $Q<0$. The corresponding distributions and profiles
$T(x)$ and $n(x)$ are shown in the leftmost panels of 
Fig. \ref{fig:inv}, where it is seen that $T$ is decreasing.
However, increasing the speed of the injected particles
leads to a sign reversal of the current, $Q>0$ and of the 
associated gradients (right panels of 
Fig. \ref{fig:inv}). A rough estimate of the conditions
where this inversion takes place, can be obtained by
equating the kinetic temperature $T_L$ of the rightmost reservoir
with (twice) of the kinetic energies of the injected particle, 
which roughly gives $T_L\approx T_0+V^2$. As seen in
Fig. \ref{fig:inv}(a) this is in agreement with the numerical
data. The above estimate (dashed white line) indeed separate
the two regions of the parameters plane having different signs
of the energy current. On the basis of this observation, 
one may argue that the velocity $V$
can be used as a further control parameter of 
the  flux direction.

\section{Conclusions}
 
In this work we have considered properties of non-equilibrium steady-states
and transport in a one-dimensional ($1D1V$) fluid in a finite space region of length $L$ undergoing MPC dynamics and interacting with thermal walls.

For Maxwellian walls we demonstrated a crossover from a normal (kinetic) transport regime to an anomalous (hydrodynamic) one, above a characteristic 
size $L_*$. Since $L_*$ grows cubically with the collision time
$\delta t$, the anomalous regime may hardly be detected  
with the sizes typically used in simulations. So we argue
that a standard diffusive description will account for the 
observation for short enough systems.
Altogether, the results nicely
fit in the general framework observed for other models \cite{Zhao2018,Miron2019,Lepri2020}. They call for a critical
consideration of the usual assumptions made in kinetic theory,
since memory effects and long-time tail may play a major role
in predicting heat and particle transport in low-dimensions.

 A natural question is whether the same 
scenario occurs in two dimensional fluids. Here, the expectation
is that in the hydrodynamic regime anomalous transport would
be associated with a logarithmic divergence of the conductivity
with size. For the 2D MPC dynamics this is confirmed by 
equilibrium simulations \cite{DiCintio2017}. Thus in principle
a similar crossover scenario should occur, although it may 
be hard to be observed in view of such a weak logarithmic dependence.

In the second part, we considered the more general case of thermal walls
injecting particles with given average (non-thermal) velocity.
We showed, that the  kinetic and hydrodynamic regimes 
are observed confirming that transport is dominated by 
bulk correlations in both cases. For injection of fast and 
relatively cold particles, short systems fail to establish
local equilibrium and display non-Maxwellian distributions
of velocities. We also showed that the injection velocity 
may be used to control current reversal.  


To conclude, let us comment on the perspective of 
applying the model to confined plasma. This requires 
considering the effect of the self-consistent electrostatic field,
solution of the associated Poisson equation.
As already noticed in Refs.\cite{Sano2001,Sano2011}, the case of 
equal masses and charges in one dimension can be mapped in
the so-called ding-dong model \cite{Prosen92}.
The latter consists of a one dimensional array of identical harmonic oscillators (each one with frequency $\omega_p$, the plasma frequency) undergoing hard-core collisions \cite{Prosen92}.
Non-equilibrium and transport properties of such a model
have been investigated in detail \cite{Prosen92,Sano2001,Sano2011}
and normal diffusive transport has been demonstrated, 
as expected since energy remains the only conserved quantity.
One may thus argue that adding MPC dynamics will not affect
significantly such a property and that the hydrodynamic regime
will not be present. However, for finite systems the situation may 
be more subtle. 
Actually, the field introduces another relevant time scale in 
the dynamics, i.e. the 
inverse of the plasma frequency $\omega_p$, associated to the collective
oscillations and Langmuir waves. This time scale should be 
compared with the collision frequency $1/\delta t$. 
Generally speaking, if $\omega_p \delta t\ll 1$ energy transport 
will be driven mostly by collisions, and it is not a priori excluded
that hydrodynamic effects may play a role over a relevant range
of scales. Those issues deserve investigation 
and will be considered in the future.

\begin{acknowledgments}
This work has been carried out within the framework of the EUROfusion consortium and has received funding from the Euratom research and training programme 2014-2018 and 2019-2020 under grant agreement No 633053. The views and opinions expressed herein do not necessarily reflect those of the European Commission.
This work is part of the Eurofusion Enabling Research project ENR-MFE19.CEA-06 \textit{Emissive divertor};
SL and RL acknowledge partial support from project MIUR-PRIN2017 \textit{Coarse-grained description for non-equilibrium systems and transport phenomena (CO-NEST)} n. 201798CZL. 
\end{acknowledgments}
 
%

\end{document}